\documentclass[conference]{IEEEtran}
\ifCLASSINFOpdf
\else
\fi
\usepackage{dependencies}

\begin{document}

\title{Performance Analysis of Decentralized Physical Infrastructure Networks and Centralized Clouds}

\author{
  \IEEEauthorblockN{
    {\anonymize{Jan von der Assen}\IEEEauthorrefmark{1}, 
\anonymize{Christian Killer}\IEEEauthorrefmark{1}\IEEEauthorrefmark{2}, \anonymize{Alessandro De Carli}\IEEEauthorrefmark{2} \anonymize{Burkhard Stiller}\IEEEauthorrefmark{1}}}
\IEEEauthorblockA{{\IEEEauthorrefmark{1}Communication Systems Group CSG, Department of Informatics IfI, University of Zürich UZH}\\
{\anonymize{Binzmühlestrasse 14, CH--8050 Zürich, Switzerland}} \\
{\anonymize{E-mail: [vonderassen, killer, stiller]@ifi.uzh.ch}}}
\IEEEauthorblockA{\IEEEauthorrefmark{2}\anonymize{Acurast Association, Dammstrasse 16, 6300 Zug, Switzerland, E-Mail [christian, alessandro]@acurast.com}}
}

\maketitle

\begin{abstract}
The advent of Decentralized Physical Infrastructure Networks (DePIN) represents a shift in the digital infrastructure of today's Internet. While Centralized Service Providers (CSP) monopolize cloud computing, DePINs aim to enhance data sovereignty and confidentiality and increase resilience against a single point of failure. Due to the novelty of the emerging field of DePIN, this work focuses on the potential of DePINs to disrupt traditional centralized architectures by taking advantage of the Internet of Things (IoT) devices and crypto-economic design in combination with blockchains. This combination yields \sol{}, a more distributed, resilient, and user-centric physical infrastructure deployment. Through comparative analysis with centralized systems, particularly in serverless computing contexts, this work seeks to lay the first steps in scientifically evaluating DePINs and quantitatively comparing them in terms of efficiency and effectiveness in real-world applications. The findings suggest DePINs' potential to \1 reduce trust assumptions and physically decentralized infrastructure, \2 increase efficiency and performance simultaneously while improving the computation's \3 confidentiality and verifiability.

\end{abstract}

\IEEEpeerreviewmaketitle

\section{Introduction}
\label{sec:intro}
Historically, industries such as telecommunications were centralized and dominated by a few centralized authorities. The rise of the Internet fundamentally challenged this centralization of communication technologies, initially spurring a shift toward decentralization. However, this movement has seen a rebound towards centralization, especially evident in the domain of cloud computing, where monopolization was further accelerated \eg through economies of scale~\cite{cloudonomics}.

The monopolization of cloud providers has led to centralized trust in cloud providers that govern vast silos of confidential data~\cite{cloudMonopoly2023}. Large-scale data breaches and leaks~\cite{edwards2016databreaches} highlighted these implicit trust assumptions and have shown that the future infrastructure needs to be built on less centralized dependencies.

The emergence of Decentralized Physical Infrastructure Networks (DePIN) signals a significant shift driven by the need to improve data sovereignty, reduce trust centralization, and increase resilience against single points of failure. This evolution reflects broader technological trends that demand increased confidentiality in computing~\cite{mulligan2021confidentialcomputing}. 

At the core of this shift, DePINs propel a novel paradigm that merges the Internet of Things (IoT) with blockchain technology and crypto-economic design (\ie tokenomics) to design a more distributed and resilient approach to digital infrastructure. One example is to take advantage of underutilized resources, such as upcycling of phones \cite{AcurastWhitepaper}.

The significance of DePINs extends beyond the technical domain, proposing a radical change in the way users engage with physical infrastructures. From solar panel-equipped homes to decentralized versions of Google Maps built from user-contributed data (\eg Hivemapper~\cite{hivemapper}), DePINs encapsulate a vision of the future where technology enables a more inclusive and collaborative approach to building and maintaining the very backbone of society~\cite{ChuaChen2024DePIN}. 

A key concern with DePINs is scalability and performance properties, mainly due to \1 the large number of smart devices involved, \2 and the performance with respect to additional latency introduced by interactions with blockchains~\cite{xinxin}. For that reason, empirical analysis and comparison of DePINs with centralized CSPs are crucial.

Despite the promising advantages of DePINs, the body of research dedicated to exploring these networks needs to be expanded, highlighting a compelling opportunity for scientific contribution. This work addresses the critical gaps identified in the background and related literature review, specifically the need for empirical validations and performance comparisons with centralized systems. It focuses on deploying DePINs for computing resources, particularly within a serverless computing model, to then compare them with centralized CSPs. 

This work focuses on laying the foundations for comparing DePINs quantitatively with centralized approaches. By providing a direct performance comparison with traditional centralized approaches (\eg Google Cloud Platform (GCP)), this work aims to substantiate the theoretical advantages of DePINs with empirical evidence. Exploring DePINs in compute-intensive scenarios is particularly crucial, as it promises to reveal new insights into their feasibility, efficiency, and effectiveness in real-world applications.

The remainder of this paper is organized as follows. Section~\ref{sec:rw} presents the relevant background and related work, emphasizing the novelty and challenges of DePINs. Section~\ref{sec:sol} outlines the proposed solutions and methodologies to explore the potential of DePINs in the context of serverless computing. Section~\ref{sec:eval} provides an evaluation and discussion of the findings, critically assessing the performance of DePINs against centralized alternatives. Finally, Section~\ref{sec:conc} summarizes and draws preliminary considerations.

\begin{table*}[b]
\caption{Related Literature on DePINs}
\centering
\begin{tabular}{@{}llllll@{}}
    \toprule
    \textit{\textbf{Solution}} &  \textit{\textbf{Proposition}} & \textit{\textbf{Resources}} & \textit{\textbf{Evaluation}} & \textit{\textbf{Comparison to CePIN}}
    \\ \midrule
    \cite{taxonomy}, 2023 & DePIN Taxonomy & N/A & Discussion & None \\
    \cite{rollupdepin}, 2023 & Rollup-centric Architecture & Sensing, Smart Devices & None & None \\
    \cite{protocoldepin}, 2023 & DePIN Device Verification Protocol & Generic & None & None \\
    \cite{onocoy}, 2023 & Satellite Navigation Framework & Navigation & Market Analysis & Market Analysis \\
    \cite{rl-depin}, 2024 & RL-based Staking Incentivization & Generic & Simulation (IoTeX) & None \\
    This, 2024 & Mobile Serverless Computing Network & Compute & Comparative Performance Analysis & Cloud Service Providers \\
    \midrule
\end{tabular}
\label{table:rw}
\end{table*}

\newpage

\section{Background and Related Literature}
\label{sec:rw}

At the core, DePIN networks share the conceptual property of providing a token-based incentive for the shared provisioning of resources. The resources targeted by such an economically incentivized scheme refer to a wide variety of types and deployment scenarios, including file storage, computing power, energy, communication connectivity, or dedicated services~\cite{ideasoft}. Due to the novelty of the paradigm, definitions, and subsuming of related literature, research that contributes to DePIN is still sparse. For example, a recent survey has identified a handful of solutions that fulfill the conceptual model of decentralized infrastructure networks~\cite{taxonomy}. Importantly, a taxonomy of the architectural elements found in DePIN approaches has been proposed, preparing the path for scientific analysis of such solutions. Regarding the work at hand, four key elements are identified for the discussion of DePIN solutions: \1 physical hardware needed to participate in the economy, \2 governance actions defined on real-life procedures, \3 middleware to act as a broker between physical and digital components, and \4 smart contracts executing the BC-based backbone of those networks.

The related literature on DePIN networks can be summarized into two areas. First, a small number of literary works have explicitly focused on DePIN from a research perspective. For example, based on a keyword search, the aggregating search engine \textit{Google Scholar} yields only five results to the query comprising "decentralized physical infrastructure networks" or "DePIN." Four elements were published in 2023 and one in 2024 (\cf~\tablename~\ref{table:rw}), stressing the novelty and lack of research in this area. The second area is industry solutions that advertise their proposition under the term, or that could be considered DePINs under the taxonomy of~\cite{taxonomy}. For example, \textit{Filecoin}~\cite{filecoin} could be viewed as a DePIN. However, it is not explicitly advertised as such.

From the list of publications, \tablename~\ref{table:rw} highlights that several papers have not only discussed theoretical aspects of DePIN, such as the taxonomy provided by~\cite{taxonomy}. For example, while~\cite{rollupdepin} proposed a roll-up-centric architecture for smart devices, \cite{protocoldepin}~designed a generic protocol for device verification in DePIN. Where~\cite{onocoy} focus specifically on the Global Navigation Satellite System, \cite{rl-depin} present a novel Reinforcement Learning-based (RL) incentivization scheme. 

Although these literary works concentrate either on specific problems within DePIN or dedicate themselves to a specific scenario, none present results from practical experiments. From the perspective of the study at hand, none focus on DePIN for computing resources, especially when considering a serverless service model. Thus, no comparisons are made with centralized approaches. All of these claims are substantiated by the fact that these publications are either white papers or published as non-peer-reviewed preprints. Although this is not a weakness per se, given the novelty of the paradigm, it calls for a stipulation of research in this field.

\begin{table}[H]
\caption{Related Industry Solutions}
\centering
\begin{tabular}{@{}llllll@{}}
    \toprule
    \textit{\textbf{Solution}} &  \textit{\textbf{Scope}} & \textit{\textbf{Scientific Evaluation}}
    \\ \midrule
    \textit{IoTeX}& IoT, Smart Devices & None  \\
    \textit{Helium}& Wireless Networks & None  \\
    \textit{IOTA}& IoT Data & None  \\
    \textit{Streamr}& P2P Infrastructure & None  \\
    \textit{MXC}& Wireless Networks & None  \\
    \textit{Akash}& Container Runtime & None  \\
    \textit{GEODNET}& Satellite Navigation & None  \\
    \textit{\sol}& Serverless Computing & Comparative Performance Analysis  \\

    \midrule
\end{tabular}
\label{table:rw-industry}
\end{table}

Regarding existing DePINs, notable networks include \textit{IoTeX}, \textit{Helium}, \textit{IOTA}, \textit{Streamr}, and \textit{MXC}~\cite{protocoldepin}. \textit{IoTeX} and \textit{IOTA} aim to interconnect IoT and smart devices, hence, not focusing on compute-heavy scenarios~\cite{iotex, iota}. In this aspect, \textit{Helium} and \textit{MXC}  are similar since they focus on interconnecting wireless networks~\cite{helium, mxc}. Finally, while Streamr~\cite{streamr} focuses on infrastructure provisioning using a P2P model, \textit{GEODNET} incentivizes the operation of a token-based localization system. Underpinning all of these approaches is the observation that none provides scientific evidence of their effectiveness and efficiency. This is magnified when focusing on the scenario of compute-intensive scenario in DePIN. Here, \textit{akash} is the most closely related industry-driven effort. As described, users engage with Akash by procuring computing resources from cloud providers prior to deploying Docker containers on the platform. The marketplace records on-chain data about user requests, bids, lease agreements, and settlement payments. The blockchain infrastructure employed by Akash serves as an integral component. As reported, the platform can beat the cost of centralized providers. However, no scientific publications report on the performance of the platform.

Based on the above discussion leading to the summary in \tablename~\ref{table:rw-industry}, the following limitations are observed:

\begin{enumerate}
    \item[\textbf{\textit{l1:}}] A lack of studies exploring DePIN solutions -- focusing on compute resources, explicitly investigating the applicability of DePIN for serverless deployments.
    \item[\textbf{\textit{l2:}}] A lack of studies directly comparing the performance of a DePIN with a centralized alternative.
\end{enumerate}


\section{\sol{} Decentralized Serverless Cloud}
\label{sec:sol}
Widely recognized challenges of blockchains and the Internet, in general, are \1 the centralization of trust in auxiliary systems, \2 the seamless and permissionless interoperability of fragmented ecosystems, and \3 the effectiveness and confidentiality of the execution layer. \sol~ is a Layer-1 blockchain that addresses these shortcomings with a novel decentralized and serverless approach~\cite{AcurastWhitepaper}.

\begin{figure}[H]
\centering
\includegraphics[width=0.3\textwidth]{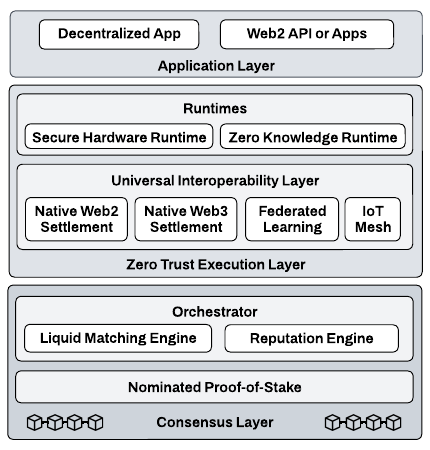}
\captionof{figure}{\sol~Architecture~\cite{AcurastWhitepaper}}
\label{fig:architecture}
\end{figure}

\sol~provides a modular separation of the consensus, execution, and application layer~(\cf~Fig.~\ref{fig:architecture}). The \sol~orchestrator is embedded in the consensus layer and contains a purpose-built reputation engine to ensure reliability and encourage honest behavior~\cite{AcurastWhitepaper}. 

So-called Consumers register their \texttt{job} on the orchestrator, offloading their computation to processors on the execution layer. Depending on the requirements, consumers can select suitable runtimes, \eg secure hardware runtimes based on smartphones' external coprocessors. 

The \sol~execution layer takes advantage of secure hardware coprocessors, removing the trust required from third parties, and reducing them to cryptographic hardness assumptions and a single root of trust (\ie secure hardware with key attestation and a coprocessor). With this approach, \sol~takes advantage of smartphone hardware and achieves efficiency and confidentiality improvements compared to CSPs~\cite{untapped}. 

\subsection{End-to-End Job Execution}
\begin{figure}[b]
\centering
\includegraphics[width=0.5\textwidth]{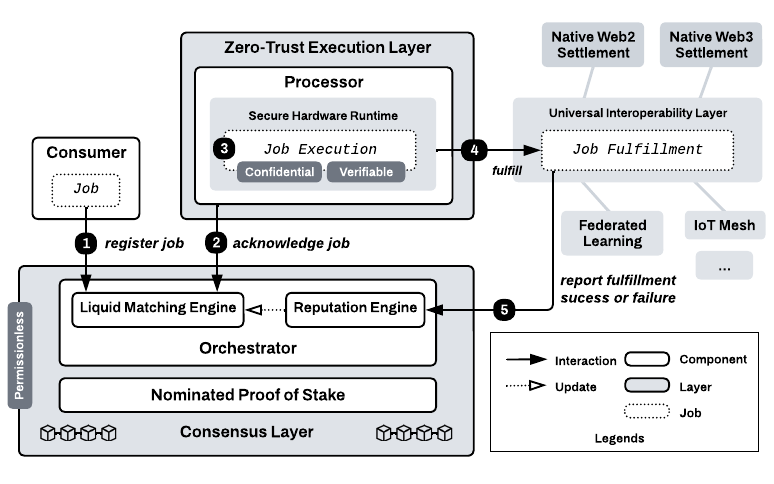}
\captionof{figure}{End-to-End Zero Trust Job Execution~\cite{AcurastWhitepaper}}
\label{fig:endtoend}
\end{figure}

The following description follows a \texttt{job} from definition and deployment to completion (\cf. Figure~\ref{fig:endtoend}). First, during \texttt{Job} Registration (1) consumers define their \texttt{job} details, \eg at what destination the \texttt{job} should be \emph{settled} \ie on which protocol the \texttt{job} \texttt{output} should be persisted (\eg Bitcoin Mainnet, Google Cloud).

During this step, the consumer must also define which processors the \texttt{job} should be executed, either \textit{(a)} on personal processors, or \textit{(b)} on selected, known processors (\eg known trusted entities), or \textit{(c)} on public processors. For \textit{(a)}, a processor reward is not required, as it is a permissioned setting. For \textit{(b)} a reward is optional, and for \textit{(c)} the liquid matching engine and the orchestrator will match processor resources with consumers' \texttt{jobs}.

In addition, more details of the \texttt{job} need to be declared, such as  \emph{scheduling} parameters, including start time, end time, the interval between executions, as well as the duration in milliseconds and the maximum start delay in milliseconds. Furthermore, specific resource management parameters, such as memory usage, network requests, and storage requirements of the \texttt{job} need to be declared. Finally, the reward for the execution of the job should be declared, as well as the minimum reputation (only applies to \textit{(c)} public processors). Then the \texttt{job} will be persisted on the Consensus Layer and reaches \texttt{OPEN} state.

Second, the \texttt{Job} is acknowledged by the processor and fetches the details from the \sol chain. Now the \texttt{job} reaches the \texttt{MATCHED} state, and no other processors will attempt to acknowledge it. 
Since \texttt{jobs} can have different scheduling configurations (\eg on demand, every minute, etc.). Therefore, if the processor acknowledges that all slots can be adhered to, the \texttt{job} reaches the \texttt{ASSIGNED} state.

Third, the \texttt{job} is executed on the processor runtime. In the illustrated example of Fig.~\ref{fig:endtoend}, the execution is performed inside of the Secure Hardware Runtime, thus \emph{confidentiality} is ascertained by secure hardware, such as an isolated and external coprocessor (\eg Google's Titan Chip~\cite{Rossi2021titan}). 

Once the \texttt{job} execution is completed, the output is delivered to the declared destination (\texttt{Job} Fulfillment (4)), which can be another Web2 system (\eg REST-API, FL model) or a Web3 system (\eg Tezos, Ethereum) that receives the output. In case of a cross-chain transaction, the processor settles the gas fees on the destination chain, since the consumer has locked the necessary reward and the gas fee amount up front when registering the \texttt{job}.

Finally, after completion, the processor reports back to the reputation engine. 
To ensure the reliability of the \sol{}~protocol, the reputation engine is continuously fed with reliability metrics, for example, right after the completion or failure of the job.

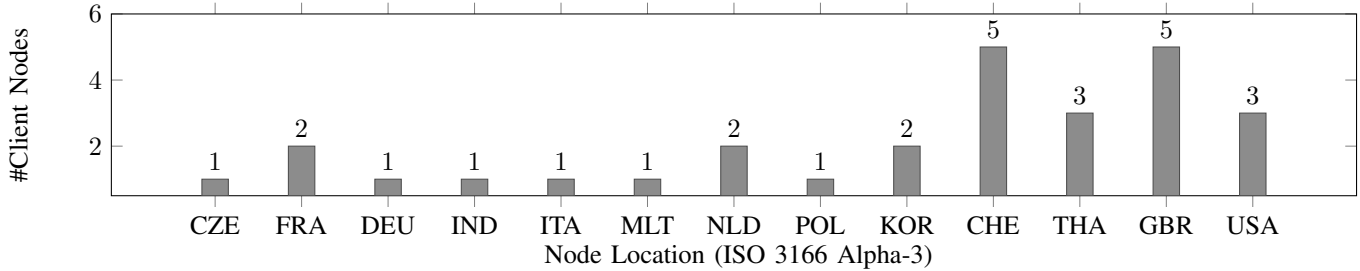
\begin{figure*}[b]

\begin{tikzpicture}  
  
\begin{axis}  
[  
    ybar,  
          cycle list={
    {fill=gray!90,draw=black!70},
    {fill=gray!10,draw=gray!40}
    },
    ymax={6},
    ylabel={\#Client Nodes}, 
    xlabel={\ Node Location (ISO 3166 Alpha-3)}, 
    width=\linewidth,
    height=4cm,
    symbolic x coords={CZE,FRA,DEU,IND,ITA,MLT,NLD,POL,KOR,CHE,THA,GBR,USA}, 
    xtick=data,  
     nodes near coords, 
    nodes near coords align={vertical},  
    ]  
\addplot coordinates {(CZE,1)(FRA,2)(DEU,1)(IND,1)(ITA,1)(MLT,1)(NLD,2)(POL,1)(KOR,2)(CHE,5)(THA,3)(GBR,5)(USA,3)};  
  
\end{axis}  
\end{tikzpicture}  
    \caption{Node Discovery}
    \label{fig:cuntries}
\end{figure*}

\section{Experiments}
\label{sec:eval}

Since DePINs are clearly a novel subject in scientific research~\cite{taxonomy} and decentralized networks are complex socioeconomic systems, a combination of empiricism~\cite{emp} and experimental analysis is needed to investigate the applicability and effectiveness of the DePIN paradigm for serverless computing. Based on the systemization described in Section~\ref{sec:sol}, the evaluations aim to investigate the following research questions.

\begin{enumerate}
    \item[\textbf{\textit{RQ1:}}] Based on the description of the system architecture of \sol{}, how many nodes have adopted the protocol and how are they distributed?
    \item[\textbf{\textit{RQ2:}}] Based on the identified set of nodes comprising the \sol{} network, how does the provided end-user service compare with centralized cloud-based ones?

\end{enumerate}

\subsection{Network and Node Discovery}\label{discovery}
To understand the deployment and adoption of the DePIN answering \textit{\textbf{RQ1}}, an experiment is performed on node discovery. Due to the lack of a centralized management node in the network that would introduce a single point of failure and the absence of a governance function for node admission (\ie that allows anyone with the appropriate hardware to participate in the permissionless network), a different approach must be employed. Since a subset of nodes participate in the serverless function provisioning service, this can be exploited to infer information about the nodes. To do so, a web-based service has been deployed on the Google Cloud Platform (GCP)~\cite{gcp}. While the RESTful endpoint does not execute any relevant load, it integrates with the logging functionality provided in the cloud service. Furthermore, a client in the \sol{} service is provisioned, which leverages the compute network to access DePIN serverless functions. More in detail, the following steps are executed as part of the experiment:
\begin{enumerate}
    \item Deploy RESTful endpoint in GCP
    \item Deploy client in the DePIN
    \item Deploy a serverless request on DePIN to fetch a resource from the RESTful endpoint.
    \item The executing node in the DePIN performs an HTTPS \texttt{GET} request on the endpoint.
    \item The cloud service collects client information (\ie the DePIN executor) including the source IPv4 or IPv6 address.
    \item The client IP addresses discovered by the cloud service are aggregated -- the above procedure is repeated at least $n$ times until no additional nodes are discovered. 
\end{enumerate}

Following said procedure, 121 function deployments have been made, and based on a service success rate of 100\%, the same number of logs were obtained in 187 seconds. After extracting the IPv4 and IPv6 addresses (\ie the DePIN nodes sending a request while executing the serverless function) from the logs, the subnets in which these addresses are announced in BGP were looked up using the BGP Looking Glass provided by~\cite{hurricaineelectric}, thus heuristically reducing the number of nodes having multiple IPv4 or IPv6 addresses from the same uplink (\ie accounting for address changes and multihoming). Then, the \textit{whois} directory service was used to obtain the geographical locations from where the prefixes are announced. Although this still has some ambiguity, each subnet was analyzed individually. For example, if an IP address belongs to a network service provider (\eg VPN providers, cloud providers), it would have been eliminated. Ultimately, the distribution of countries depicted in \figurename~\ref{fig:cuntries} presents an overview of current geographical penetration. 

\begin{figure}[H]
    \centering
\begin{tikzpicture}
\begin{axis}[
    ybar stacked,
	bar width=15pt,
	nodes near coords,
    enlargelimits=0.2,
    legend style={at={(0.65,0.90)},
      anchor=north,legend columns=-1},
    ylabel={\#Devices},
    symbolic x coords={Pixel, Samsung, Moto, TCL, Xiaomi, Redmi, Oneplus, Lenovo, Oppo},
    xtick=data,
    x tick label style={rotate=45,anchor=east},
    height=5cm,
    width=8cm,
      cycle list={
    {fill=gray!70,draw=black!70},
    {fill=gray!10,draw=gray!40}
    }
    ]
\addplot+[ybar] plot coordinates {(Pixel,10) (Samsung,20) (Moto, 5) (TCL, 2) (Xiaomi, 1) (Redmi, 3) (Oneplus, 1) (Lenovo, 1) (Oppo, 1)};
\addplot+[ybar] plot coordinates {(Pixel,17) (Samsung,1) (Moto, 0) (TCL, 0) (Xiaomi, 0) (Redmi, 0) (Oneplus, 0) (Lenovo, 0) (Oppo, 0)};
\legend{\strut Models, \strut Devices }
\end{axis}
\end{tikzpicture}
    \caption{Devices Employed in the \sol{} DePIN}
    \label{fig:devicesuseragent}
\end{figure}
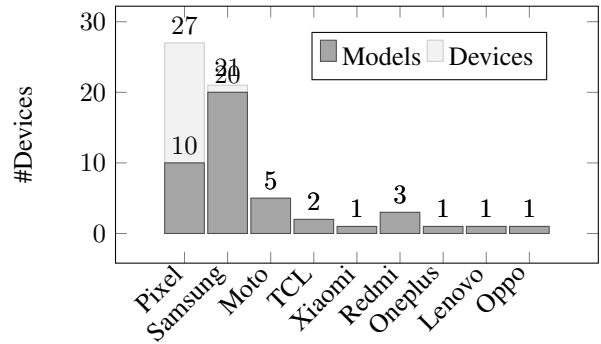

To answer \textit{\textbf{RQ1}}, 30 nodes have been discovered in 14 countries, effectively covering three continents (\ie Europe, Asia, and North America). Although this analysis is not exhaustive (\ie presents only a snapshot of the deployed nodes), a longitudinal study could offer additional insight while requiring one to account for additional pitfalls (\eg address changes, multihoming). Furthermore, all function deployments have been successful, indicating that there were no Sybil nodes at the time of the experiment. However, due to the limited complexity of the workload required by the serverless function, additional experiments are needed to investigate the behavior of the nodes.

Additional information on \textbf{\textit{RQ1}} involves an analysis of the nodes themselves. Here, the experiment relies on the \texttt{User-Agent} header set by the clients performing the requests on the cloud instance. Although even client IP addresses do not inherently correctly identify the geographic location (\eg they could be tunneled), relying on the user agent information inherently implies that the client information is trusted. However, with this trust assumption in mind, the following conversations are made. First, all nodes report running the Dalvik virtual machine, as expected based on the system description in Section~\ref{sec:sol}. Second, 62 distinct devices are discovered. Relating this to the 30 unique IP addresses discovered implies that nodes employ $\approx$2 devices per node. Ten nodes operated \sol{} on top of Android 11 (\ie 16\%), 13 operated Android 12 (\ie 20\%), 27 operated Android 13 (\ie 44\%), and 12 devices used Android 14 (\ie 19\%). As indicated in \figurename~\ref{fig:devicesuseragent}, the 62 devices comprise 43 different models across nine smartphone vendors.

\subsection{Comparative Performance Analysis}
In the second experiment, a computationally expensive workload is deployed to the following instances on the respective platforms, which enable serverless computation. In \tablename~\ref{tab:platforms}, each platform is characterized in terms of deployment location and runtime environment. For each platform, the workload depicted in Algorithm~\ref{alg:sieve} is deployed. The algorithm uses the Sieve of Eratosthenes to efficiently identify and generate prime numbers up to a specified maximum value (\ie \texttt{max}). It initializes arrays to track whether the numbers are marked as composite (sieve) and to store prime numbers (primes). The algorithm iterates through each number, marking its multiples as a composite in the sieve array and adding the unmarked numbers to the prime array. By systematically eliminating multiples, the algorithm efficiently identifies prime numbers within the given range, providing an optimized method for generating a list of primes up to the specified maximum value. The algorithm's complexity is expressed as $O(n \log \log n)$.

\setlength{\tabcolsep}{3pt}

\begin{table}[H]
\caption{Platforms Involved in Comparative Analysis}
\centering
\begin{tabular}{@{}llllll@{}}
    \toprule
    \textit{\textbf{Platform}} &  \textit{\textbf{Location}} & \textit{\textbf{Runtime}} & \textit{\textbf{Hardware}}
    \\ \midrule
    \sol{} & Distributed & V8 ECMA-262  &  Randomly Sampled \\
    \textit{Local1} & Local & Node.js v18  & i7-8650U@1.90GHz, 16GB \\
    \textit{Local2} & Local & Node.js v18  & 7502P@2.5GHz, 128GB \\
    \textit{Microsoft Azure}& Iowa, USA & Node.js v18 & Unknown \\
    \textit{Amazon AWS}& Sweden & Node.js v18 & Unknown  \\
    \textit{Google Cloud}& Iowa, USA & Node.js v18 & Unknown  \\
    \midrule
\end{tabular}
\label{tab:platforms}
\end{table}

\begin{algorithm}
\caption{Benchmark Algorithm -- Sieve of Eratosthenes}
\label{alg:sieve}
\begin{algorithmic}[1]
\Function{getPrimes}{$max$}
    \State $sieve \gets []$
    \State $i, j \gets 0$
    \State $primes \gets []$
    \For{$i \gets 2$ to $max$}
        \If{not $sieve[i]$}
            \State \Call{push}{$primes, i$}
            \For{$j \gets i \times 2$ to $max$ by $i$}
                \State $sieve[j] \gets$ \textbf{true}
            \EndFor
        \EndIf
    \EndFor
    \State \Return $primes$
\EndFunction
\end{algorithmic}
\end{algorithm}

A JavaScript adapter was written and deployed to each platform to execute the functions. Then, a test runner executes each function several times with \texttt{max=50,000,000}, measuring the completion time of the function. This value was selected since it was the largest one successfully completed across all platforms.


%
%
%
%
%

\begin{figure}[b]
    \centering
    \includegraphics[width=\linewidth, trim={.3cm .2cm 1.3cm 1.4cm},clip]{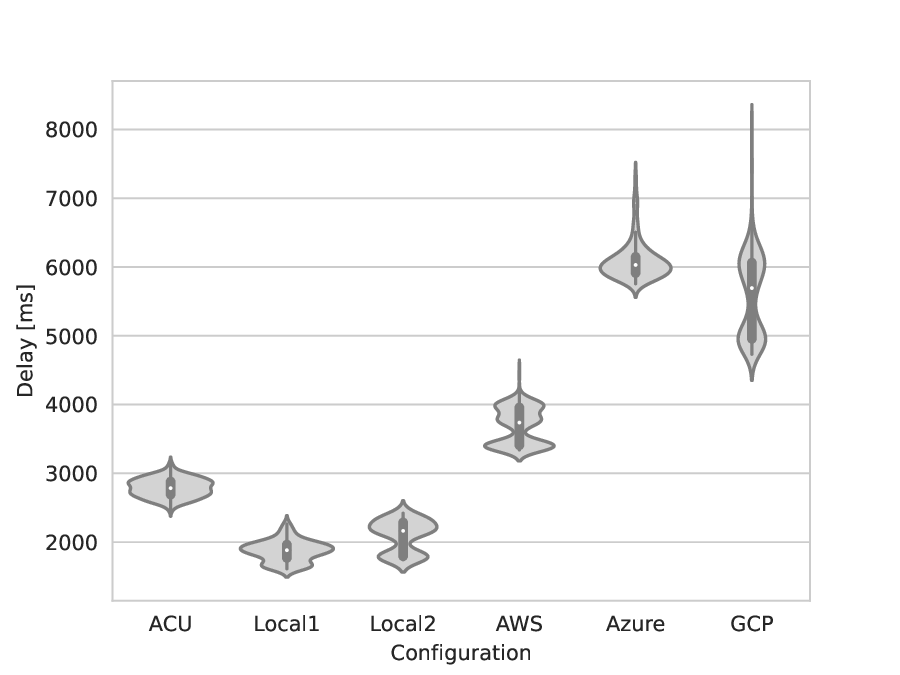}
    \caption{CPU-intensive Benchmark~\cite{data}}
    \label{fig:violin-delays}
\end{figure}

In \figurename~\ref{fig:violin-delays}, the distribution across all response times are plotted for each platform, respectively, based on the analysis of $\approx$1000 measurement samples. When comparing the average delay, it can be seen that the \sol{} platform completes the previously described workload in 2790 ms, while AWS requires 3683 ms and Google Cloud 5565 ms. Finally, Azure exhibits the highest delay with an average response time of 6102 ms. Thus, it can be observed that the globally decentralized network provided by the DePIN may indeed be a viable alternative for computationally intensive scenarios. Furthermore, it can be seen that the response times distribution through \sol{} is narrowly distributed (\ie average of 2790 ms, with a standard deviation of $\approx$134). However, it is supported by a heterogeneous and ungoverned physical infrastructure.

Answering \textbf{\textit{RQ2}} by comparing the performance of \sol{} with two centralized and local approaches (\ie not serverless), two interesting observations arise. As depicted in \figurename~\ref{fig:violin-delays}, both local configurations (\ie \textit{Local1} and \textit{Local2}) outperform \sol{} by a comparatively small margin. Compared to the mobile CPU on \textit{Local1}, \sol{} presents an increase of 30.55\% and an increase of 46.04\% compared to the server. However, when considering the ad hoc deployment model of \sol{} (\ie without hardware or software maintenance required for a client), the solution presents a strong improvement compared to centralized solutions. However, given the early and growing stage of \sol{} (\cf Sec.~\ref{discovery}), further experiments would be required to evaluate scalability.

\subsection{Power Efficiency}
Aside from computational effectiveness, power efficiency is vital. Hence, performance in relation to power consumption must be considered. The previously described workload mainly benefits from single-threaded performance since I/O can be largely neglected. Using the server from \textit{Local2} as an illustrative scenario, the workload completely saturates one CPU core for an average duration of 2092 ms. As described by the manufacturer, the CPU requires 180 Watts of power~\cite{hpepyc}. Ignoring the consumption of peripherals, each core would require 5.625 watts. Thus, a single run of the benchmark algorithm would require $3.268750 \times 10^{-3}$ Wh. In comparison, one Cortex-A715 core leveraged by \sol{} requires less than 300 mW. When factoring in the extended computation time, one iteration would require $2.275833\times 10^{-4}$ Wh, marking a drastic decrease in performance per watt.

In other words, with a single watt-hour, the above workload can be executed $\approx$4394 times using~\sol{} or $\approx$306 in a local scenario. Furthermore, it should be reiterated that this comparison draws on the values obtained in a local environment. Although cloud providers stress the importance of power efficiency~\cite{awsgreenblabla}, the authors have not identified any sources presenting actual numbers of the cloud provider's efficiency of serverless functions.

\section{Summary and Preliminary Considerations}
\label{sec:conc}
Due to the novelty of the DePIN paradigm, this paper identified a lack of studies exploring DePIN solutions focusing on the provisioning of compute resources, especially when a serverless deployment model is considered. Furthermore, an overarching lack of direct comparisons between existing DePIN approaches and their centralized alternatives.

To address these limitations, this paper introduced the architecture of \sol{}. Within this use case, the present work investigated the deployment of nodes within this DePIN and analyzed their service offering in terms of computational effectiveness and power efficiency. Furthermore, these results were contrasted with the results obtained from Cloud Service Providers (CSP). These initial results suggest that a mobile DePIN can perform comparatively to a local execution environment and potentially outperform remote CSPs, especially in terms of power efficiency.

However, in addition to acknowledging the limitations of the study (\ie the limited number of devices and the reliance on client-side data), future research will investigate the performance of additional workloads and consider the security of the platform against several threats.

\section*{Acknowledgment}
This work has been partially supported by \textit{(a)} the \anonymize{Acurast Association} and \textit{(b)} the \anonymize{University of Zürich UZH}.

\balance
\bibliographystyle{IEEEtran}
\bibliography{assets/main.bib}
\noindent \small{\\All links above were last accessed on April 12th, 2024.}

\end{document}